\documentclass[prb,twocolumn,amsmath,amssymb,floatfix]{revtex4}
\usepackage{graphicx}
\usepackage{bm}
\usepackage{bbold} 

\usepackage{color}
\usepackage{soul}
\usepackage{ulem}

\def\gae{\lower 2pt \hbox{$\, \buildrel {\scriptstyle >}\over {\scriptstyle \sim}\,$}}
\def\lae{\lower 2pt \hbox{$\, \buildrel {\scriptstyle <}\over {\scriptstyle \sim}\,$}}

\newcommand{\bra}{\langle}
\newcommand{\ket}{\rangle}

\newcommand{\eref}[1]{Eq.~\ref{#1}}
\newcommand{\sref}[1]{Sec.~\ref{#1}}
\newcommand{\fref}[1]{Fig.~\ref{#1}}

\newcommand{\aref}[1]{App.~\ref{#1}}
\newcommand{\rcite}[1]{Ref.~\citenum{#1}}

\newcommand{\be}{\begin{equation}}
\newcommand{\ee}{\end{equation}}
\newcommand{\beq}{\begin{eqnarray}}
\newcommand{\eeq}{\end{eqnarray}}
\newcommand{\nn}{\nonumber}
\newcommand{\init}{\mathrm{init}}

\begin{document}

\title{Dynamic trapping near a quantum critical point}
\author{Michael Kolodrubetz, Emanuel Katz, Anatoli Polkovnikov}
\affiliation{Physics Department, Boston University, 590 Commonwealth Ave., Boston, MA 02215, USA}

\begin{abstract}
The study of dynamics in closed quantum systems has been revitalized by the emergence of experimental systems that are well-isolated from their environment.  In this paper, we consider the closed-system dynamics of an archetypal model: spins driven across a second order quantum critical point, which are traditionally described by the Kibble-Zurek mechanism.  Imbuing the driving field with Newtonian dynamics, we find that the full closed system exhibits a robust new phenomenon -- dynamic critical trapping -- in which the system is self-trapped near the critical point due to efficient absorption of field kinetic energy by heating the quantum spins. We quantify limits in which this phenomenon can be observed and generalize these results by developing a Kibble-Zurek scaling theory that incorporates the dynamic field. Our findings can potentially be interesting in the context of early universe physics, where the role of the driving field is played by the inflaton or a modulus field.
\end{abstract}

\maketitle

The Kibble-Zurek mechanism describes the behavior of systems ramped slowly across a continuous phase transition\cite{Kibble1980_1, Zurek1985_1}. It has been difficult to observe experimentally due in part to non-universal dynamics that often dominate late in the ramp and overwhelm the universal critical dynamics, although many attempts have been made (see \rcite{delCampo2014_1} for a recent review). Recent work has extended the phenomenology of Kibble-Zurek to a full non-equilibrium scaling theory in the vicinity of the critical point \cite{Deng2008_1,DeGrandi2011_1,Chandran2012_1,KolodrubetzPRL2012_1}, which has the advantage of yielding robust universal behavior out of equilibrium. However, its observation requires the often challenging task of tuning and measuring the system very close to its critical point.

In this paper, we will consider a simple extension to the Kibble-Zurek mechanism of a quantum phase transition, in which we treat the field $\lambda$ that drives a system across its quantum critical point as a dynamical ``particle'' in its own right.  With this simple change, we find a new phenomenon in which the initial kinetic energy of the $\lambda$ degree of freedom is efficiently absorbed by heating of the quantum critical degrees of freedom, trapping $\lambda$ at or near the critical point. By using the power of Kibble-Zurek scaling theory, we can very generally predict when such trapping will occur solely in terms of the equilibrium critical exponents of the quantum critical point. This gives universal non-equilibrium dynamics without fine-tuning, which may prove useful in experiment studies of Kibble-Zurek scaling. Also, as we will later argue, this dynamic trapping may have connections to inflationary physics.

\begin{figure}
\includegraphics[width=.7\linewidth]{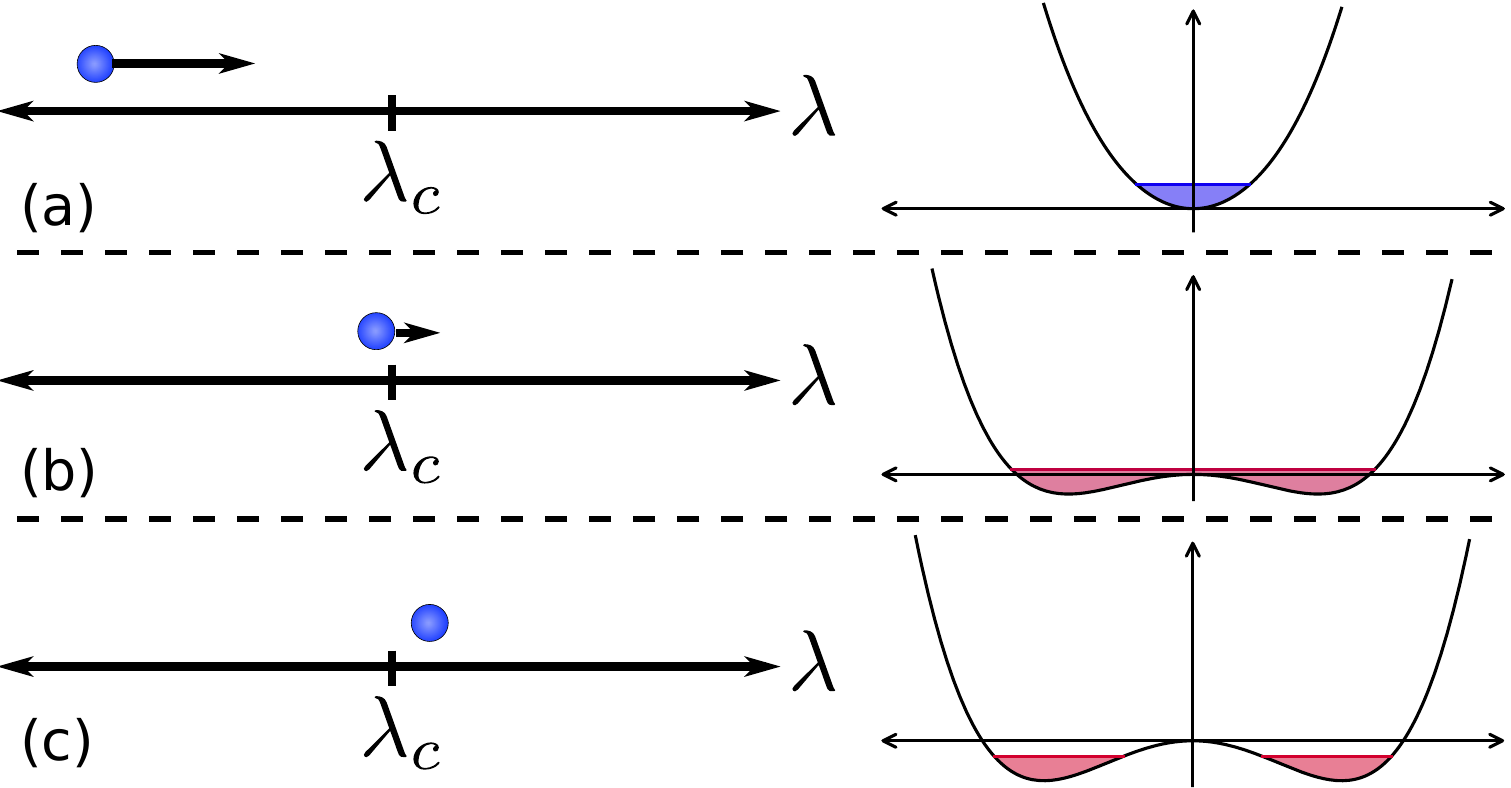}
\caption{Basic idea of the critical trapping phenomenon. 
The control field $\lambda$ is initialized in the disordered phase with initial velocity toward the QCP. (b) As the quantum degrees of freedom (e.g., spins) heat up, the dynamic field slows down, until (c)  $\lambda$ can get trapped at or near the critical point.}
\label{fig:critical_trapping_basics}
\end{figure}

More concretely, consider a generic Hamiltonian $H_0(\lambda, \phi({\bf x}))$ in $d$ spatial dimensions which can be statically tuned by $\lambda$ across a second order quantum critical point at $\lambda_c$. Here $\phi({\bf x})$ represent the quantum degrees of freedom in the system, which for notational simplicity we refer to as spins\footnote{We note that our arguments hold for arbitrary fields undergoing a second order quantum phase transition, not just spin systems.}.  We assume that $\lambda$ is macroscopic and thus described by classical Newtonian dynamics with some bare mass $M_\lambda$ and bare external potential $V(\lambda)$. 
The Hamiltonian of the full system is
\be
H = H_0(\lambda, \phi({\bf x})) + \frac{p_\lambda^2}{2 M_\lambda} + V(\lambda)~.
\label{eq:H_tot_generic}
\ee

Together the system and control parameter are isolated, and the total energy is conserved. The parameter $\lambda$ can represent either an external degree of freedom, such as a  macroscopic object that is coupled to the system\cite{Berry1989_1, Dalessio2014_1} or an internal (e.g., mean-field) degree of freedom such as a superconducting gap\cite{Barankov2004_1} or the effective mass in a large $N$ field theory\cite{Chandran2013_1}. Here we focus on an external degree of freedom that drives the system across a second order phase transition.
We assume that the spins are initialized in the ground state at some $\lambda_\init$ far from the critical point and that $\lambda$ has some initial velocity $v_\init$ toward the critical point. For an externally driven field (i.e., $M_\lambda =\infty$) the system's response is described by the well-known Kibble-Zurek (KZ) mechanism~\cite{Kibble1980_1, Zurek1985_1}, which predicts universal non-adiabatic dynamics characterized by an emergent length scale~\cite{Zurek1985_1, Damski2005_1, Zurek2005_1, Polkovnikov2005_1, Dziarmaga2005_1, Chandran2012_1}. In particular, external ramping across the critical point at velocity $v_\init$ should lead to heating of the spins at the critical point that scales as $Q \sim L^d v_\init^{(d+z)\nu\over 1+\nu z}$, where $\nu$ and $z$ are the equilibrium correlation length and dynamic critical exponents, respectively~\cite{Polkovnikov2011_1}.

If we now allow feedback of the spins on the dynamics of $\lambda$ ($M_\lambda \neq \infty$), we can approximate the fate of the system via energy conservation. On the one hand, if the spins remained in their ground state, then the kinetic energy $K_c$ of $\lambda$ at the critical point would be
\be
K_c = \frac{M_\lambda v_\init^2}{2}+\big[V(\lambda_c)-V(\lambda_\init)\big]+\big[E_\mathrm{gs}(\lambda_c)-E_\mathrm{gs}(\lambda_\init)\big],
\label{eq:energy_cons_K}
\ee
where $E_\mathrm{gs}(\lambda)$ is the ground state energy of the spin system. This dissipationless limit defines the bare velocity $v_c$ upon reaching the critical point
\be
K_c  = \frac{1}{2} M_\lambda v_c^2 ~.
\ee
On the other hand, we have seen that the energy $Q_c$ absorbed by the spin system scales as
\be
Q_c \sim L^d v_c^{(d+z)\nu\over 1+\nu z}~.
\label{eq:Q_c}
\ee
We expect that the parameter will be trapped if the energy the spins want to absorb is greater than the initial kinetic energy:
\be
Q_c > K_c \implies \mu v_c^{\frac{1}{1+\nu z} \left[ 2 + \nu (z - d) \right]} \lae 1 ~,
\label{eq:energy_cons_QK}
\ee
where $\mu = M_\lambda / L^d$ is the mass density of the $\lambda$ field~\footnote{Because the parameter $\lambda$ is extensive in the system size its mass has to scale with the volume of the system.}. This equation has very interesting implications. In low dimensions, where the exponent in \eref{eq:energy_cons_QK} is positive:
\be
\frac{1}{1+\nu z} \left[ 2 + \nu (z - d) \right]>0 ~~\Longleftrightarrow~~ d<z+{2\over \nu} \equiv d^\ast ~,
\label{eq:d_c}
\ee
the parameter is always trapped below a certain threshold velocity. However, in high dimensions $d>z +2/\nu$, there is no trapping at small velocities and $\lambda$ can freely pass through the critical point. For example, in standard Ginzburg-Landau type theories with $z=1$, $\nu$ saturates at $1/2$ above $d=3$, yielding a critical dimension $d^\ast = 5$ below which trapping will occur.

To justify these considerations we will analyze a specific exactly solvable model -- the transverse-field Ising (TFI) chain in $d=1$ spatial dimension with a dynamical transverse field:
\be
H_0 = - \sum_j \big[ (1 - \lambda) s^z_j + s^x_j s^x_{j+1} \big] ~,
\label{eq:H_tfi}
\ee
where $s$ are the Pauli matrices. The TFI chain undergoes a quantum phase transition at $\lambda_c=0$ from a disordered paramagnet ($\lambda < 0$) to $\mathbb Z_2$ symmetry-broken ferromagnet ($\lambda > 0$) with exponents $\nu=z=1$, yielding trapping if $\mu v_c \lae 1$.  Because the TFI chain has an explicit UV cutoff, our previous arguments require that the trapping velocity is sufficiently small to give $K_c$ and $Q_c$ much smaller than the cutoff; in general systems, we similarly require that excitations caused by the dynamics occur at a scale well below the scale of the leading irrelevant operator to ensure the validity of the critical field theory.

\begin{figure}
\includegraphics[width=\linewidth]{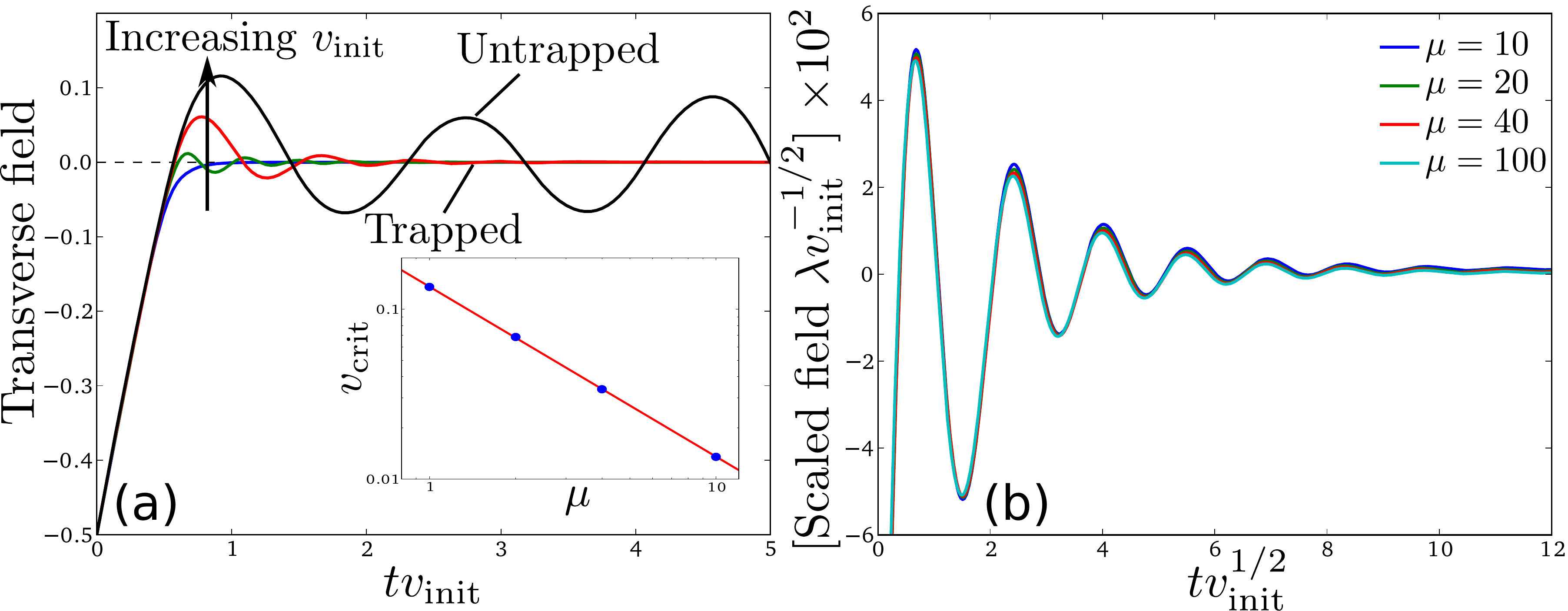}
\caption{Demonstration of critical trapping in the TFI model for $V(\lambda) = -E_\mathrm{gs}(\lambda)$, such that $v_c = v_\init$ (see main text). (a) As $v_\init$ is increased from 0.02 to 0.06, 0.1, and 0.14 at fixed $\mu=1$, the field undergoes a trapping/untrapping transition. (inset) The critical value of $v_\init$ for trapping (blue dots) scales as $1/\mu$ (red line) as predicted from a KZ analysis. (b) Scaling collapse of the dynamics at fixed $\mu v_\init = 0.06$.}
\label{fig:tfi_transition_no_pot}
\end{figure}

The TFI chain is integrable; it can be solved by a Jordan-Wigner transform from spin 1/2's to spinless fermions to yield a quadratic Hamiltonian~\cite{Sachdev1999_1}.  We then numerically simulate the exact coupled spin and field equations of motion for large system sizes ($L \gae 10000$), which are checked for system size independence to ensure convergence to the thermodynamic limit. Details of our simulations can be found in \aref{app:simulation_details}. As we will see, neither the macroscopic dynamics of the $\lambda$ field nor the KZ scaling are sensitive to the integrability of the theory~\cite{KolodrubetzPRL2012_1}, so we expect that the results we present will be generic.

We first carry out the simulations in a potential $V(\lambda)$ chosen to cancel out the ground state energy: $V(\lambda)=-E_\mathrm{gs}(\lambda)$. This ensures that there is no force on $\lambda$ when the spins remain in their ground state, and thus $v_c = v_\init$ in \eref{eq:energy_cons_K}. Later we consider a more general setup without such compensation. We start the system in its ground state at large negative $\lambda$ and give it positive initial velocity so that it heads toward the critical point. With this choice of potential the equations of motion are
\beq
\nn M_\lambda \ddot\lambda &=&-\langle \psi(t) |\partial_\lambda H_0 |\psi(t)\rangle +\partial_\lambda E_\mathrm{gs}(\lambda)
\\ i \frac{d |\psi(t)\ket}{dt} & = & H_0\big( \lambda(t) \big) |\psi(t) \ket ~,
\label{eq:newton_tfi}
\eeq
where $|\psi(t) \ket$ is the spin wave function. We find that the resulting behavior of the magnetic field is in perfect agreement with our qualitative considerations. For a fixed value of the mass density $\mu$, our data show a transition in the long time behavior of $\lambda$ as the velocity is increased past a critical threshold (\fref{fig:tfi_transition_no_pot}a). Examining this trapping/untrapping transition for a range of $\mu$, we see that the prediction of \eref{eq:energy_cons_QK} is born out: the transition happens at a constant value of the initial momentum density $\mu v_\init=\mu v_c$ (\fref{fig:tfi_transition_no_pot}a, inset). Furthermore, postulating that $\mu v_c$ is the only important scale in the problem, we see in \fref{fig:tfi_transition_no_pot}b that the entire dynamics of the field undergoes scaling collapse if the mass is varied at fixed $\mu v_c$.

\fref{fig:tfi_transition_no_pot}b foreshadows the existence of a non-equilibrium scaling theory that is a natural extension of the conventional Kibble-Zurek mechanism in the presence of a dynamic field.  We show in \aref{sec:supp_kz_scaling} that the equations of motion for $\lambda$ and the spin wave function are consistent with a Kibble-Zurek scaling theory with characteristic time and length scales given by scaling dimensions $[\lambda]=1/\nu,\; [t]=-z,\; [v_c]=[\dot\lambda]=1/\nu+z$:
\be
\ell_{KZ}=v_c^{-\nu/(1+\nu z)} ~,~ t_{KZ} = v_c^{-\nu z/(1+\nu z)}~,~\lambda_{KZ} = v_c^{1/(1+\nu z)}
\label{eq:kz_scaling_vinit}
\ee
The only difference from conventional KZ scaling is that externally-imposed ramp rate is replaced by the initial velocity. In particular, it is clear that a scaling solution is possible if both sides of \eref{eq:newton_tfi} have the same scaling dimensions, implying that
\[
[\mu]+[\lambda]+2 z-d=z-[\lambda] ~\Longleftrightarrow ~ [\mu]=d-z-2/\nu ~.
\]
Combining this with \eref{eq:kz_scaling_vinit}, we see that a scaling solution is possible if the mass $\mu$ scales as
\be
\mu_{KZ}=v_c^{\nu(d-z)-2\over 1+\nu z} ~,
\label{eq:mu_KZ}
\ee
which matches with the prediction of \eref{eq:energy_cons_QK} for the trapping/untrapping transition.  Therefore, the dynamics of the trapped field, as well as the spin observables, should be universal at fixed ratio $\mu/\mu_{KZ}$ as shown numerically in \fref{fig:tfi_transition_no_pot}b. We note in passing that \eref{eq:mu_KZ} gives a scaling dimension of mass that is consistent with that of the mass renormalization of $\lambda$ found elsewhere via adiabatic perturbation theory\cite{Dalessio2014_1}. Thus the trapping condition in \eref{eq:d_c} is equivalent to a negative scaling dimension of the mass renormalization, i.e., divergence of the dressed mass of $\lambda$ at the critical point.

Now consider a more generic situation where the external potential does not exactly compensate the ground state energy. Because there is no general principle of choosing such a potential \footnote{One might argue that the ground state energy itself is a natural potential to choose (i.e., $V(\lambda)=0$).  However, different models in the same universality class need not have the same ground state potentials, and scaling theory does not predict the behavior of the ground state energy in the vicinity of the critical point.}, we simply expand it as a Taylor series near the critical point. Generically the leading term will be linear, but if the critical point has additional symmetries, the leading term can be higher order. Therefore we consider the case $V(\lambda) = -E_\mathrm{gs}(\lambda)+\alpha L^d \lambda^r$, where $\alpha$ is the strength of an arbitrary power law potential added to the flat potential. This modifies the equation of motion to
\be  
M_\lambda\ddot \lambda =-\langle \psi(t) |\partial_\lambda H_0 |\psi(t)\rangle + \partial_\lambda E_\mathrm{gs} -r \alpha L^d \lambda^{r-1}~.
\label{eq:eom_with_pot}
\ee
The scaling dimension of the force due to the power law potential is clearly $[L^d \lambda^{r-1}]=-d+(r-1)/\nu$, while the scaling dimension of the other term is $[\partial_\lambda H]=z-1/\nu$. We see that this additional force is relevant provided that
\be
-d+\frac{r-1}{\nu}<z-{1\over \nu} \; \Leftrightarrow \; r<\nu (z+d).
\ee
For our case this implies $r<2$, so a constant force -- corresponding to $r=1$ -- is clearly relevant. 

Naively one would expect that this constant force will simply destroy the localization and the system will fall downhill through the critical point. However, the situation is much more interesting if the coefficient $\alpha$ is small.  Let us then revisit our previous analysis for small $\alpha$ and consider the specific case of starting at rest ($v_\init = 0$) with non-zero $\lambda_\init$. We had argued in the absence of a potential that the system will stop as long as the velocity $v_c$ is smaller than a critical value (see \eref{eq:energy_cons_QK}), setting an upper bound on $v_c$.  This bound was derived with the implicit assumption that the time scale $t_c$ to reach the critical point was much larger than the inverse of the initial gap, $\Delta_\init^{-1} \sim \lambda_\init^{-\nu z}$, so that the initial dynamics is adiabatic. Let's first consider this regime, which translates to $\lambda_\init \gae (\alpha / \mu)^{1/(1+ 2\nu z)}$ for a particle starting from rest. In this regime $v_c \sim \sqrt{\lambda_\init \alpha / \mu}$, and demanding that $Q/L^d \gae \alpha \lambda_\init$ implies trapping for 
\be
 \lambda_\init ~\lae ~\frac{1}{\alpha} \left(\frac{1}{\mu}\right)^{\frac{\nu(d+z)}{2-\nu(d-z)}}~.
\label{eq:trapping_init_adiab}
\ee 

Now consider $\lambda_\init \lae (\alpha / \mu)^{1/(1 + 2\nu z)}$, in which case $t_c$ is much shorter than gap time scale and the dynamics of $\lambda$ approach those of an instantaneous quench. In this regime the renormalized mass of $\lambda$ is much larger than the bare mass\cite{Dalessio2014_1}, so the bare mass term becomes completely irrelevant to the long time dynamics (see \aref{sec:supp_kz_scaling}).  Consequently, $\lambda_\init$ is effectively the only scale, and we have that $Q \sim \lambda_\init^{\nu(d+z)}$. This implies trapping for $Q \gae K\sim \alpha \lambda_\init \Leftrightarrow \lambda_\init \gae \alpha^{1/[\nu(d+z)-1]}$.  Combined with \eref{eq:trapping_init_adiab}, we therefore expect trapping if
\be
\alpha^{\frac{1}{\nu(d+z)-1}} ~\lae ~\lambda_\init ~\lae ~\frac{1}{\alpha} \left(\frac{1}{\mu}\right)^{\frac{\nu(d+z)}{2-\nu(d-z)}}~,
\label{eq:trapping_ineq}
\ee 
which for small $\alpha$ and $d> 1 / \nu - z$ yields a non-trivial region.  One expectation from \eref{eq:trapping_ineq} is that starting directly at the QCP ($\lambda_\init=0$) should not lead to trapping, which we readily confirm numerically. Another prediction is that above some critical value $\alpha > \alpha_c \sim \mu^{[\nu(d+z)-1]/[\nu(d-z)-2]}$, \eref{eq:trapping_ineq} cannot be satisfied and thus no trapping will occur for any initial conditions.

\begin{figure}
\includegraphics[width=\linewidth]{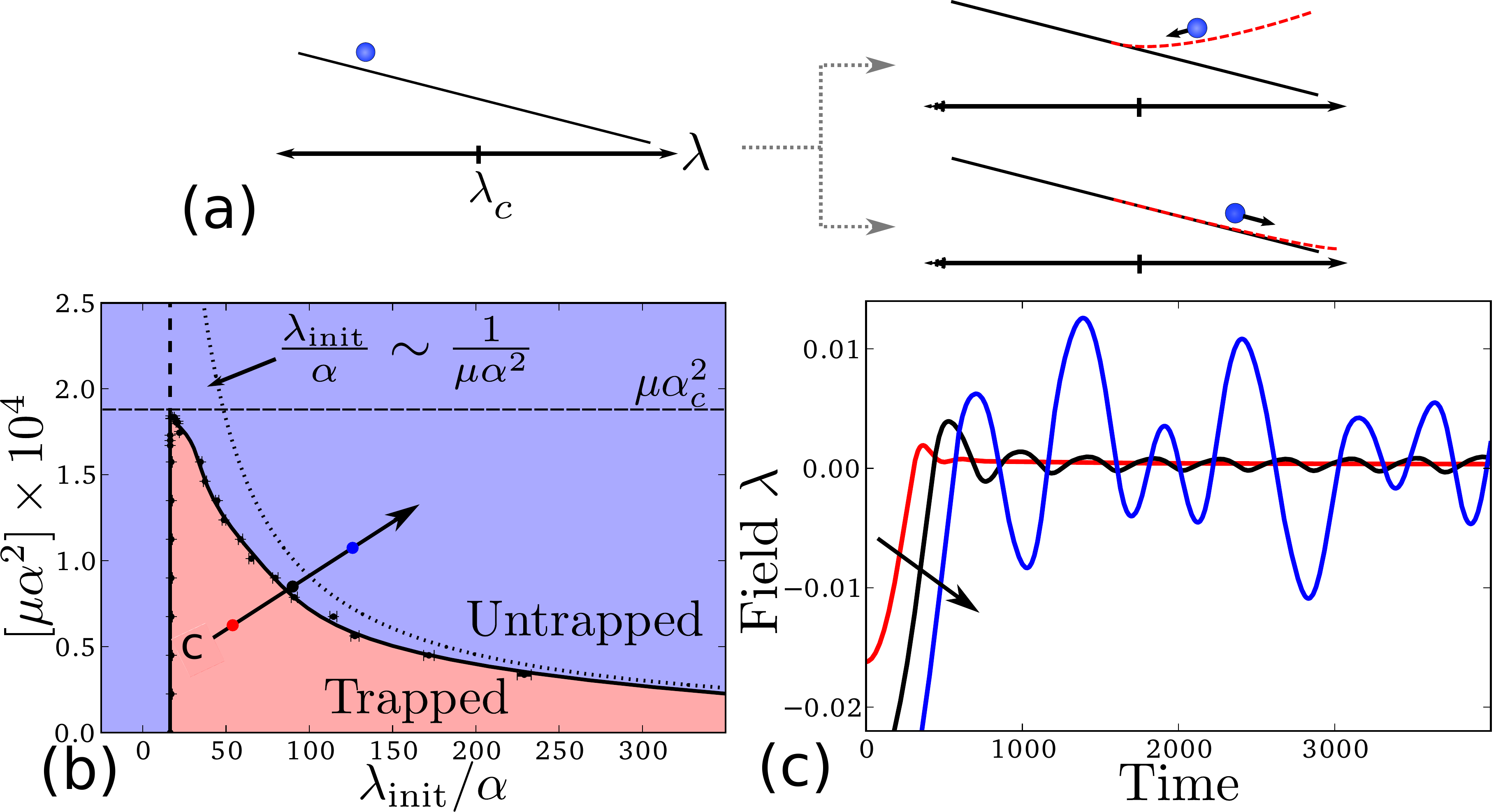}
\caption{Trapping in the presence of an external potential. (a) Starting from rest, as the particle falls through the critical point, excitations act as an effective potential (red dashed line) drawing the system toward the QCP. Depending on the strength of this dressing, the system can be either trapped (upper) or untrapped (lower). Scaling arguments suggest that trapping should occur for $1 \lae \lambda_\init / \alpha \lae 1/(\mu \alpha^2)$, which is encompassed in the phase diagram shown in (b). To visualize the transition, we show a one-dimensional cut across the phase boundary in (c), showing a field which is trapped (red), untrapped (blue), and near the transition (black). Data in (b) and (c) is shown for $\alpha=-3\times 10^{-4}$, which numerically appears to approach the $\alpha\to 0$ scaling limit.}
\label{fig:tfi_transition_linear_pot}
\end{figure}

Substituting the exponents for the TFI chain we have that trapping occurs for
\be
1 ~\lae~ \frac{\lambda_\init}{\alpha} ~\lae~ \frac{1}{\mu \alpha^2} ~.
\label{eq:trap_top_tfi}
\ee
More rigorously, we expect a phase diagram in which the trapping transition is a universal function of scaling variables $\mu \alpha^2$ and $\lambda_\init / \alpha$ as shown in \fref{fig:tfi_transition_linear_pot}. The phase diagram demonstrates our prediction of a maximum slope $\alpha_c \sim 1/\sqrt{\mu}$ above which no trapping occurs. Furthermore, the phase boundaries from \eref{eq:trapping_ineq} describe low $\mu \alpha^2$ well due to a clean separation between initially adiabatic and diabatic regimes in this limit.

All of the above arguments are exact in the scaling limit $\alpha \to 0$, but for realistic systems the approximations must break down at long times, where RG-irrelevant operators can affect the dynamics. An important example of this in the TFI chain are weak integrability-breaking interactions, which at long times are expected to lead to thermalization \cite{Deutsch1991_1, Srednicki1994_1, Rigol2008_1, StarkArxiv2013_1}. For our results to remain robust against interactions, we must see whether a thermal state at the same energy is trapped, i.e., if there exists a local maximum of the entropy as $\lambda$ is varied. We analyze this thermodynamic trapping for the TFI chain in \aref{app:rg_irr_sc_terms} and find thermodynamic trapping for initial energies $\lambda_\init / \alpha \gae 15.3$. This thermodynamic trapping region encompasses the entire dynamically trapped region, so we expect that our results will be robust against interactions. Finally we note that in some cases which we call untrapped (e.g., \fref{fig:tfi_transition_no_pot}a and \fref{fig:tfi_transition_linear_pot}c), the field quasiperiodically oscillates around the critical point rather than escaping to infinity. At long times, this behavior could be modified by irrelevant operators as well and it is possible that these operators could damp the quasiperiodic oscillations and thus increase the trapped region.

One crucial aspect of all of our results is their robustness, which is inherited from the universality of Kibble-Zurek scaling theory. For example, consider the Hamiltonian of a complex interacting scalar field $\phi$ living in $d$ spatial dimensions:
\be
H_\phi = \int d^d x \big[ |\Pi(x)|^2 + |\nabla \phi(x)|^2 + \lambda |\phi(x)|^2 + u |\phi(x)|^4 \big] ~,
\label{eq:H_phi4}
\ee
where $\Pi$ is the momentum conjugate to $\phi$.  As $\lambda$ is driven from large positive to large negative values, the system undergoes a phase transition from disordered ($\bra \phi \ket = 0$) to ordered, where $\bra \phi \ket \neq 0$ spontaneously breaks the $U(1)$ symmetry.  This type of phase transition occurs in a wide range of systems, from the Ginzburg-Landau theory of superconductivity \cite{Tinkham2004_1} to the dynamics of the Higgs field in particle physics. This system has standard Ginzburg-Landau critical exponents so, as argued earlier, this model should be trapped at low $v_\init$ for $d < 5$.

If we think of \eref{eq:H_phi4} as describing the Higgs field in the early universe, it is conceivable that the mass term $\lambda$ is related to the value of the inflaton field, a field whose potential energy has been proposed to drive expansion, or a modulus field as in \rcite{Kofman2004_1}.  Setting aside the complications due to expansion for future work, here we focus on the simpler case of a field whose energy is extensive in a system of constant volume. If we initialize the system on the disordered side of the transition with a non-zero velocity of the $\lambda$ field directing it toward the quantum critical point, then for slow enough initial velocities (or weak enough slope $\alpha$), we have seen that the $\lambda$ field should be trapped near the critical point, which corresponds to zero Higgs mass. Such a situation can occur naturally in the slow roll inflation paradigm (see \cite{Baumann2009_1} for a review). If these arguments hold up against the additional complications of inflation, then this is potentially an interesting picture of how the Higgs field can be ``trapped'' at small mass in the absence of fine-tuning.

\begin{figure}
\includegraphics[width=.7\linewidth]{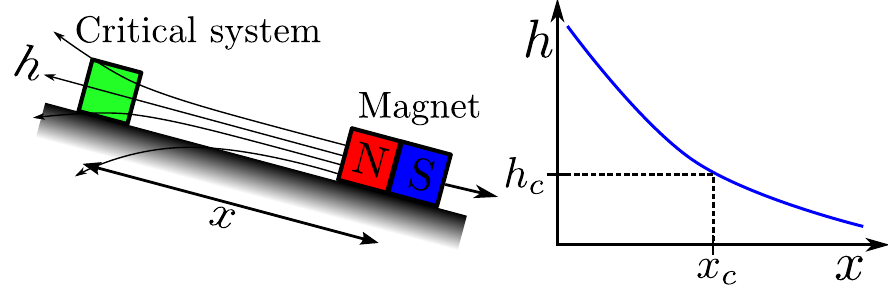}
\caption{A simple experiment to illustrate critical trapping. As a permanent magnet moves down the hill (position $x$), its field $h(x)$ drives a stationary quantum critical system through its phase transition. If the system parameters are in the trapped region of the phase diagram (\fref{fig:tfi_transition_linear_pot}b), then the magnet is trapped very near this critical point ($x(t\to\infty) \approx x_c$).}
\label{fig:crude_experimental_example}
\end{figure}

In a condensed matter context, there are a number of scenarios where we would expect this physics to be relevant. A simple example is illustrated in \fref{fig:crude_experimental_example}. The idea is that a stationary quantum critical system (e.g., a ferromagnet or a superconductor) is coupled to the magnetic field generated by a moving magnet at position $x$. As the magnet slides down a hill, the changing magnetic field experienced by quantum critical system drives it across a phase transition. If the slope of the hill is sufficiently small, where ``sufficiently small'' is encapsulated by the phase diagram in \fref{fig:tfi_transition_linear_pot}b, then the permanent magnet is trapped very close to this critical point. This dynamical phenomenon can yield some counter-intuitive results. For instance, if we consider a superconductor whose phase transition is driven by having the magnetic field $h$ drop below the superconducting critical field, then a standard undergraduate physics experiment says that the Meissner effect will start to repel the magnet (or for a different geometry, levitate it). But our dynamical trapping phenomenon would have the opposite effect, leading to a magnet that is actually pulled \emph{toward} the superconductor just at the onset of superconductivity. In realistic settings one must account for finite temperature dissipative effects that will eventually destroy this trapping, which will be the subject of a future paper. However, if one is at low enough temperatures within the ``quantum critical fan'' \cite{Sachdev1999_1}, then for a certain time scale this counter-intuitive dynamical trapping should have a significant effect before thermal excitations eventually untrap it.

A similar story holds for other coupled systems and may be particularly relevant for cases where a single system has two different types of degrees of freedom. In the solid state one could imagine scenarios where a ferromagnetic field from stationary electrons drives the superconducting transition of itinerant electrons or where the magnetic field from a spin impurity drives a quantum phase transition in its bath. These ideas are particularly relevant in ultracold atom experiments, where hybrid sytems combining multiple atomic species \cite{Lewenstein2007_1}, atoms in higher Rydberg states \cite{Pohl2009_1}, and cold atoms coupled to cold ions \cite{Bissbort2013_1} have become prevalent. One use of the atom/ion system is to simulate isolated condensed matter systems, where the ions have long-range Coulomb interactions that allow them to self organize into a lattice. Future  experiments may be able to use this to drive a quantum phase transition, such as the bosonic superfluid to Mott insulator transition that would occur upon increasing the ionic lattice spacing to decrease tunneling. If one were to do a ramp or quench of some parameter $\lambda$ that controls the bosonic hopping (e.g. the strength of the ion trap or the ion/atom interaction strength), then the ionic lattice could become dynamical trapped near the value where the transition occurs. The key difference from the experiments in which the Mott insulator to superfluid transition has been seen in optical lattices \cite{Greiner2002_1} is that the degrees of freedom driving the transition are a dynamical part of the system. Therefore the strong backaction of the quantum critical system plays a huge role in the global dynamics.

In conclusion, we have used intuition drawn from Kibble-Zurek scaling theory to find a novel trapping/untrapping transition of a dynamic field occurs in systems near their quantum critical point, which is seen to occur without fine-tuning for a wide range of initial conditions. This idea readily generalizes to classical phase transitions which also exhibit the critical slowing down at the heart of the dynamic trapping mechanism. We expect this idea to have applications for a wide variety of systems, from condensed matter and cold atom systems that are well isolated from their environment to possibly inflationary scenarios.  Also, the thermodynamic trapping mechanism discussed above can potentially have implications as far-reaching as high-temperature superconductors and other systems where the order parameter (playing the role of $\lambda$) is often observed to be ``trapped'' in the vicinity of a hypothesized hidden critical point \cite{Park2006_1,Shibauchi2014_1}. An interesting open question is the robustness of these results against quantum fluctuations of the $\lambda$ field. Berry has semi-classically argued that such a quantum field experiences an additional correction to the potential that is proportional to the quantum metric tensor \cite{Berry1989_1}, which diverges at the critical point. While this term is suppressed in the thermodynamic limit for an extensive mass $M_\lambda$, it may prove an important correction at long times for large but finite systems. 

\emph{Acknowledgements -- } We would like to acknowledge funding from NSF DMR-1206410 and PHY-1211284 as well as AFOSR FA9550-13-1-0039.
EK was supported by DOE grant DEFG02-01ER-40676.  We would also like to thank Luca d'Alessio, Yariv Kafri, Anushya Chandran, Sid Parameswaran, and Shivaji Sondhi for many useful discussions.


\bibliography{../../References/References}

\appendix 

\section{Simulating the dynamic TFI chain}
\label{app:simulation_details}

Throughout this paper, we illustrated our ideas using the quintessential example of a quantum phase transition: the one-dimensional transverse-field Ising chain \cite{Sachdev1999_1}. As a reminder, the spin Hamiltonian of the model is 
\be
H_0 = - \sum_j \left[  (1-\lambda) s^z_j + s^x_j s^x_{j+1} \right]~,
\label{eq:H_tfi_supp}
\ee
where $\lambda$ is the transverse field and $s^i_j$ are Pauli matrices on site $j$.  We consider a TFI chain with $L$ sites and periodic boundary conditions.  To this spin Hamiltonian, we add classical dynamics to the transverse field and an energy offset to remove the ground state energy $E_\mathrm{gs}(\lambda)$, yielding the full system Hamiltonian
\be
H = H_0(\lambda) + \frac{1}{2}  \big( \mu  L \big) \dot \lambda^2 - E_\mathrm{gs}(\lambda) ~.
\label{eq:H_field_supp}
\ee
We ignore quantum fluctuations of the field $\lambda$ because it is extensive, and therefore fluctuations vanish in the thermodynamic limit (TDL).

As derived in the main text, the equations of motion for the field are given by 
\be
\ddot \lambda = - \frac{ \bra \partial_\lambda H_0 / L \ket - \bra \partial_\lambda H_0 / L \ket_0}{\mu} ~,
\label{eq:field_dyn_cons_supp}
\ee
which for the TFI chain is given by $\partial_\lambda H_0 = \sum_j s^z_j$.  
Meanwhile, the quantum evolution is tractable for large system sizes because the TFI chain is integrable. Starting from the Hamiltonian in \eref{eq:H_tfi_supp}, we can diagonalize it by doing a Jordan-Wigner transformation from spins to fermions, followed by a Fourier transform (cf. \rcite{Sachdev1999_1}).  Implicit in this transformation are the parity of the number of sites ($L$) and of the total number of spin-up particles ($N_\uparrow$), which is a conserved quantity in the TFI model.  For simplicity, we choose both of these quantities to be even, which amounts to considering an even number of fermions with anti-periodic boundary conditions  \cite{Dziarmaga2005_1}; in the thermodynamic limit, this assumption will not be important.  Then the Hamiltonian is separable, $H_0=\sum_{k>0} H_k$, where $k=2\pi(n+1/2)/L$ for $n=0, 1, \ldots, L/2-1$ are the positive momenta in the first Brillouin zone and, following the conventions of \rcite{KolodrubetzPRL2012_1},
\beq
\nonumber H_k &=& (1-\lambda-\cos k)(c_k^\dagger c_k+c_{-k}^\dagger c_{-k}-1) \\
&&+ \sin k (c_k^\dagger c_{-k}^\dagger + c_{-k} c_k) ~.
\label{eq:H_k_supp}
\eeq
This quadratic Hamiltonian respects conservation of both momentum and fermion parity, and thus can only excite momenta $+k$ and $-k$ in pairs.  One can easily show that the parity in each momentum sector is even by adiabatic continuation from $\lambda = \pm \infty$, so since we start from the ground state, we can safely neglect the unpaired-momentum sector.  Thus, each $H_k$ reduces to a $2\times 2$ Hilbert space, which we can rewrite in terms of pseudo-spin operators $\sigma_k$, in which the $(+k,-k)$ pair is filled if $\sigma^z_k=1$ and empty if $\sigma^z_k=-1$.  The pseudo-spin Hamiltonian is 
\be
H_k = (1-\lambda - \cos k )\sigma^z_k + (\sin k )\sigma^x_k ~.
\label{eq:H_k_sigma_supp}
\ee

The spins begin in their ground state $|\psi\ket = \bigotimes_k |\psi_k\ket$, which is a product state over momentum sectors.  Each sector evolves independently under the Schr\"odinger equation 
\be
i \frac{d |\psi_k \ket}{dt} = H_k |\psi_k \ket ~,
\label{eq:schrodinger_spins}
\ee
where $H_k$ is a function of the instantaneous value of $\lambda$ (\eref{eq:H_k_sigma_supp}).  Similarly, given the wave function $|\psi\ket$, the value $\bra s^z_\mathrm{tot} \ket = 2 \sum_k \bra \sigma^z_{k>0} \ket$ can be easily computed, where the sum is over positive momenta because the fermions are excited in $\pm k$ pairs.  Therefore, 
we simulate the coupled dynamics of the field and the spins by a Suzuki-Trotter decomposition \cite{Trotter1959_1,Suzuki1976_1}, first evolving the field via \eref{eq:field_dyn_cons_supp} for a small time step $\Delta t$ with fixed $|\psi\ket$, then evolving $|\psi \ket$ with fixed field.  We reach the continuous-time thermodynamic limit by making $\Delta t$ smaller and $L$ larger until no further changes in the observables can be seen.

\section{Effect of integrability-preserving irrelevant perturbations}
\label{app:irrel_pert}

\begin{figure}
\includegraphics[width=\linewidth]{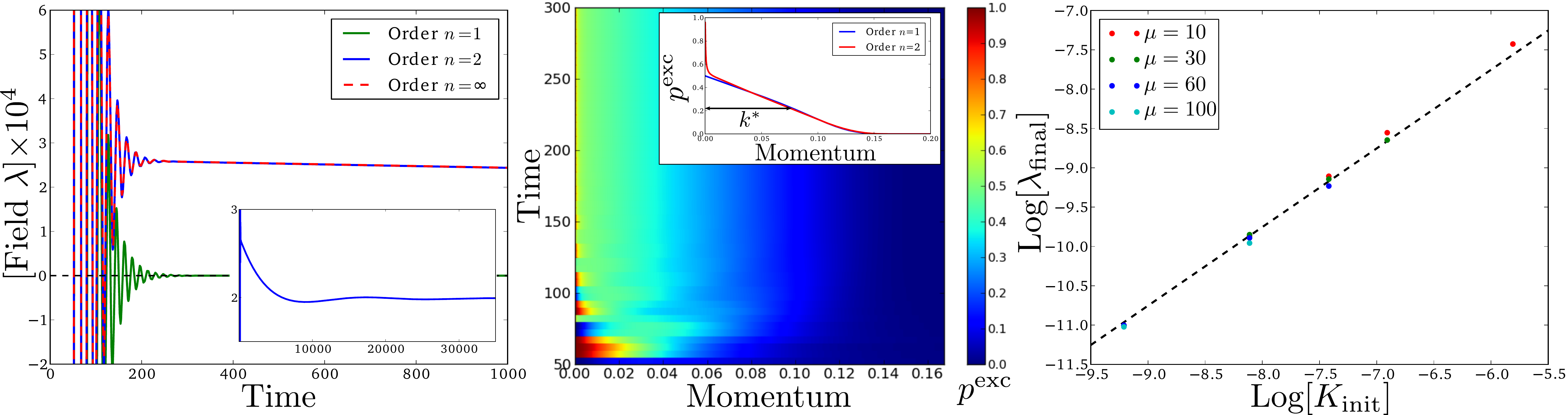}
\caption{Higher order corrections to the scaling theory.  (a) Long-time behavior of the transverse field, where the dispersion relations are truncated at $n$-th order in momentum.  The Hamiltonians are given by \eref{eq:H_k_linear} for $n=1$ (linear), \eref{eq:H_k_quadr} for $n=2$ (quadratic) and \eref{eq:H_k_sigma_supp} for $n=\infty$ (untruncated). The inset shows even longer times for $n=\infty$. (b) Excitation probability $p^\mathrm{exc}$ as a function of momentum for various times during the ramp truncated to linear order. The inset compares $p^\mathrm{exc}$ for $n=1$ and $n=2$ truncation at late time $t=1000$. The characteristic momentum scale $k^*$ is labeled. (c) Log-log plot of initial kinetic energy versus $\lambda$ at late times, which is extrapolated from data similar to panel (a). The data shows consistency with the expected scaling $\lambda_\mathrm{final} \sim K_\mathrm{init}$ (dashed line). Data in panels (a) and (b) is shown for $\mu=20$ and $v_\init=10^{-2}$.}
\label{fig:supp_fig1}
\end{figure}

When we examine the dynamics of the full TFI chain with a dynamic field in detail, we find that for even for small $v_\init$, at late times the field does not settle directly to the QCP (see \fref{fig:supp_fig1}a).  Zooming into these dynamics shows that the field appears to undergo damped oscillations on two separate time-scales before finally settling to a value 
$\lambda_\mathrm{final} \neq 0$. Based on the Kibble-Zurek arguments that will be described in more detail in the next section, we expect that in the limit $v_\init \to 0$, the field will settle at $\lambda_\mathrm{final}=0$ if $\mu v_\init$ is small enough for the field to get trapped.  Therefore, we expect that the non-zero value of $\lambda_\mathrm{final}$ seen in \fref{fig:supp_fig1} should be a result of irrelevant operators that vanish in the Kibble-Zurek limit.

To see this, we expand the sine and cosine functions in \eref{eq:H_k_sigma_supp} around $k=0$ to $n$-th order.  At leading order ($n=1$), the Hamiltonian becomes
\be
H_k^{(1)} = -\lambda \sigma^z_k + k \sigma^x_k ~.
\label{eq:H_k_linear}
\ee
We refer to this linearized case as the scaling theory, which will be justified in the next section.
As seen in \fref{fig:supp_fig1}a, this linearized Hamiltonian settles exactly to the QCP at late times.  At second order, we find
\be
H_k^{(2)} = -\left( \lambda + \frac{k^2}{2} \right) \sigma^z_k + k \sigma^x_k ~,
\label{eq:H_k_quadr}
\ee
and similarly at higher orders.  For the field dynamics in \fref{fig:supp_fig1}, we clearly see that this second-order approximation is sufficient to describe the offset of $\lambda_\mathrm{final}$ from zero. 

The scaling of this offset can be easily understood from analyzing the late-time generalized Gibbs ensemble (GGE) \cite{Rigol2007_1}, the density matrix obtained by dephasing this integrable system at late times.  The GGE is determined by the field $\lambda$ and the conserved excitation probabilities $p^\mathrm{exc}(k)$.  To see this, we can rewrite the mode Hamiltonian as
\beq
\nn H_k &=& -\epsilon_k \big[ \sigma^z_k \cos \theta_k + \sigma^x_k \sin \theta_k \big]
\\ \nn \epsilon_k &=& \sqrt{(\lambda + \cos k - 1 )^2 + \sin^2 k}
\\ \tan \theta_k &=& \frac{\sin k}{\lambda + \cos k - 1} ~.
\eeq
The ground state at momentum $k$ is a Bloch vector aligned parallel to the effective magnetic field
angle $\theta_k$, while the excited state is anti-parallel to this field.  Therefore, 
\be
\bra \sigma^z_k \ket_\mathrm{gs}
= \cos \theta_k  = \frac{\lambda + \cos k - 1}{\epsilon_k}  = -\bra \sigma^z_k \ket_\mathrm{es} ~.
\label{eq:sigma_z_gs}
\ee
The force on $\lambda$ is proportional to $\sum_k ( \bra \sigma^z_k \ket - \bra \sigma^z_k \ket_\mathrm{gs})$,
which can be written in the simple form
\be
\sum_k ( \bra \sigma^z_k \ket - \bra \sigma^z_k \ket_\mathrm{gs}) = -2 \sum_k p^\mathrm{exc}_k \bra \sigma^z_k \ket_\mathrm{gs} ~.
\ee

The excitation probability $p^\mathrm{exc}$ is a strictly positive function, an example of which is shown in \fref{fig:supp_fig1}b. If we empirically approximate it by $p^\mathrm{exc}_k \approx \frac{1}{2} e^{-k / k^*}$ with some characteristic momentum scale $k^*$, then from energy conservation in the linearized approximation, we find that
\beq
\nonumber Q / L &=& \frac{1}{2\pi} \int_0^{\pi} dk \epsilon_k p^\mathrm{exc}_k \approx
\frac{1}{4\pi} \int_0^{\pi} dk k e^{-k / k^*}
\\ & \sim &  (k^*)^2 \sim K_\init / L ~.
\eeq
Meanwhile, in the linearized approximation, $\bra \sigma^z_k \ket_\mathrm{gs} \to \lambda / \sqrt{\lambda^2 + k^2}$, meaning that $\mathrm{sgn} \bra \sigma^z_k \ket_\mathrm{gs} = \mathrm{sgn} \lambda$.  This causes oscillations around $\lambda=0$ and, in particular, ensures that $\lambda=0$ is the only fixed point of the field evolution in the linearized approximation.  

However, if we include second order corrections to $\bra \sigma^z_k \ket_\mathrm{gs}$, the sign of this expectation value -- and thus its contribution to the acceleration $\ddot \lambda$ -- will change at some non-zero value of $k$.  From a second-order expansion of the numerator in \eref{eq:sigma_z_gs}, it is clear that this sign change occurs at $\lambda = k^2 / 2$.  Then, assuming that $k^*$ is the only momentum scale in the problem, we would predict that the final value of $\lambda$ would scale as $\lambda^* \sim (k^*)^2 \sim \mu v_\init^2$.  This prediction is confirmed in \fref{fig:supp_fig1}c, justifying our assumption that the second order expansion provides a good description of the late-time dynamics. It is clear from this analysis that the second order and higher terms in the expansion become irrelevant in the KZ limit ($v_\init \to 0$).

\section{Kibble-Zurek scaling with a dynamic field}
\label{sec:supp_kz_scaling}

\begin{figure}
\includegraphics[width=.4\linewidth]{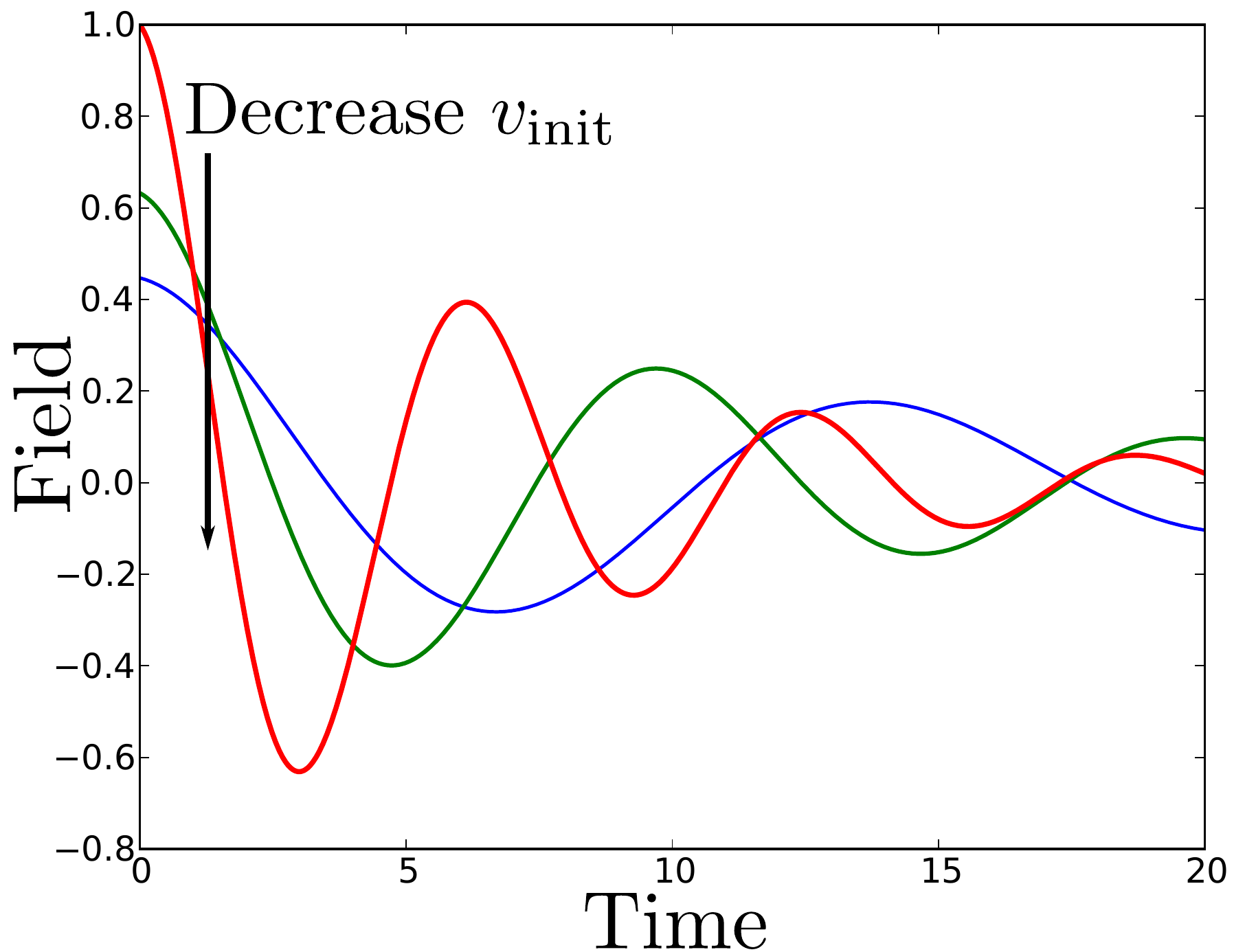}
\caption{Illustration of KZ scaling of arbitrary field profiles.  The KZ limit is taken by decreasing the initial velocity while rescaling the time and field axes by $t_{KZ}$ and $\lambda_{KZ}$ as given in \eref{eq:kz_scaling}.}
\label{fig:supp_fig2}
\end{figure}

To understand many of the results in the previous section, we can generalize the arguments in, e.g., \rcite{KolodrubetzPRL2012_1} to show KZ scaling in the presence of a dynamic field.  Before dealing with the dynamics of the field, we generalize the Kibble-Zurek scaling arguments in Ref. \citenum{KolodrubetzPRL2012_1} to an arbitrary ramping protocol $\lambda(t)$ such that the spins are in their ground state at time $t=0$.  Without loss of generality, assume that there is a non-zero initial velocity $v_\init=\dot \lambda (0)$.  We claim that a useful way of rescaling this protocol to take the Kibble-Zurek limit of $v_\init \to 0$ is to define the family of protocols,
\be
\lambda_{v_\init}(t) = \lambda_{KZ} \tilde \lambda (\tilde t = t / t_{KZ}) ~,
\label{eq:gen_ramp_protocol}
\ee
parameterized by the initial velocity.  Here, $\lambda_{KZ}$ and $t_{KZ}$ are the standard KZ scales for $v_\init$ \cite{Polkovnikov2011_1}:
\be
\lambda_{KZ} \equiv v_\init^{1/(1+\nu z)}~,~ t_{KZ} \equiv v_\init^{-\nu z/(1+\nu z)} ~,
\label{eq:kz_scaling}
\ee
and $\tilde \lambda(\tilde t)$ is an arbitrary continuous function defined on the range $t \geq 0$, with $\frac{d \tilde \lambda}{d \tilde t} \big|_{\tilde t=0} = 1$ (see \fref{fig:supp_fig2}). It is then easy to see that the initial velocity of the protocol $\lambda_{v_\init}(t)$ is $v_\init$.

The dynamics near an isolated quantum critical point becomes universal upon taking the limit $v_\init \to 0$ of the above family of protocols.  Note that this is precisely the standard Kibble-Zurek scaling form with $v_\init$ playing the role of the
fixed velocity \cite{Chandran2012_1}; indeed one can recover standard Kibble-Zurek scaling using the linear protocol $\tilde \lambda(\tilde t) = \tilde \lambda_\init + \tilde t$. However, \eref{eq:gen_ramp_protocol} can be generalized to arbitrary ramps, and in particular, we will show it can be used to self-consistently give scaling dynamics with a dynamic field.

To continue the analogy with standard Kibble-Zurek scaling, we wish to define a set of rescaled parameters, which we denote by adding a tilde:
\beq
\nn \tilde \lambda = \lambda / \lambda_{KZ} &\equiv& \lambda v_\init^{-1/(1+\nu z)} 
\\ \nn \tilde t = t / t_{KZ} &\equiv& t v_\init^{\nu z/(1+\nu z)} 
\\ \nn \tilde k = k \ell_{KZ} &\equiv& k v_\init^{-\nu/(1+\nu z)} 
\\ \tilde \mu = \mu / \mu_{KZ} &\equiv& \mu v_\init^{(\nu z+2-d\nu)/(1+\nu z)} ~,
\eeq
where $k$ is the momentum.  Note that for the case of the TFI chain, $\tilde \mu = \mu v_\init$ is the initial momentum, which we now show plays the role of a scaling parameter in the dynamics.  Beginning with the mode-evolution equation
\be
i \frac{d |\psi_k\ket}{dt} = [(1-\lambda(t)-\cos k)\sigma^z_k + (\sin k) \sigma^x_k]
|\psi_k\ket ~,
\ee
we can expand to third order in the momentum and rewrite the dynamics in terms of the scaling variables:
\beq
\nonumber i \frac{d |\psi_{\tilde k}\ket}{d\tilde t} &=& \Big[ (-\tilde \lambda(\tilde t)+
\frac{{\tilde k}^2}{2} v_\init^{1/2} + O(\tilde k^4)) \sigma^z_{\tilde k} ~+~
\\ &&~ (\tilde k - 
\frac{{\tilde k}^3}{6} v_\init  + O(\tilde k^3)) \sigma^x_{\tilde k}\Big] |\psi_{\tilde k}\ket ~.
\eeq
In the limit $v_\init \to 0$, all but the leading order terms vanish, and the wave function becomes a function of just $\tilde t$ and $\tilde k$.  This justifies our choice to refer to the linearized mode Hamiltonian (\eref{eq:H_k_linear}) as the scaling limit.

The scaling relations apply not only to the wave functions, but to expectation values of certain observables.  For instance, the operator $s^z_\mathrm{avg} = \frac{1}{L} \sum_j s^z_j$ should have scaling dimensions of inverse length, such that we predict a scaling form
\be
\bra s^z_\mathrm{avg}(t) \ket - \bra s^z_\mathrm{avg}(t) \ket_0= \ell_{KZ}^{-1} f(\tilde t)~,
\label{eq:supp_szavg_scaling}
\ee
for some universal function $f$, where $\bra \cdots \ket_0$ is the instantaneous ground state expectation value.  We see that this form emerges by substituting the mode wave function $|\psi_{\tilde k}(\tilde t)\ket$ and taking the TDL:
\beq
\nn \bra s^z_\mathrm{avg}\ket & = & \frac{1}{L} \sum_j \bra s^z_j \ket = 
\frac{1}{L} \sum_k \bra s^z_k \ket 
\\ \nn &=&\frac{1}{L} \frac{L}{2\pi} \int dk \bra \psi_{\tilde k} (\tilde t) | 
s^z_k | \psi_{\tilde k} (\tilde t) \ket
\\ &=&v_\init^{1/2} \underbrace{\frac{1}{2\pi} \int d \tilde k \bra \psi_{\tilde k} (\tilde t) | 
s^z_{\tilde k} | \psi_{\tilde k} (\tilde t) \ket}_{f(\tilde t)} ~,
\eeq
and similarly for the ground state expectation value. This type of scaling form should hold for a wide variety of operators \cite{Deng2008_1,Chandran2012_1,KolodrubetzPRL2012_1}; as our scaling postulate, we assume this to hold for the remainder of the paper.

We now demonstrate that the equations of motion governing the dynamics of the field are consistently solved by a protocol of the same form as \eref{eq:gen_ramp_protocol}, with $\tilde \mu$ now appearing as an extra parameter:
\be
\lambda_{\mu}(t) = \lambda_{KZ} \tilde \lambda_{\tilde \mu} (\tilde t = t / t_{KZ}) ~.
\label{eq:dyn_ramp_protocol}
\ee
As before, we note that the equations of motion for $\lambda$ are 
\be
\ddot \lambda = -\frac{\bra \partial_\lambda H_0 / L^d \ket - \bra \partial_\lambda H_0 / L^d \ket_0}{\mu} ~.
\label{eq:eom_lambda_gen}
\ee
Let's assume that these dynamics yield $\lambda$ of the form in \eref{eq:dyn_ramp_protocol}. Then, by the previous arguments (i.e., a generalized form of \eref{eq:supp_szavg_scaling}), the expectation value will have the scaling form
\be
\frac{\bra \partial_\lambda H_0 \ket-\bra \partial_\lambda H_0 \ket_0}{L^d} = \frac{1}{t_{KZ} \ell_{KZ}^d \lambda_{KZ}} f_{\tilde \mu} (\tilde t) ~.
\ee
Substituting this expression into \eref{eq:eom_lambda_gen} and properly inserting powers of $v_\init$, we see that the equations of motion take the scale invariant form
\be
\frac{d^2 \tilde \lambda}{d\tilde t^2} = - \frac{1}{\tilde \mu} f_{\tilde \mu} (\tilde t) ~.
\ee
This establishes the consistency of the field motion with the K-Z scaling ansatz of the spins, so the entire dynamics is universal.

\section{Scaling with a linear potential}
\label{sec:supp_scaling_linear}

\begin{figure}
\includegraphics[width=\linewidth]{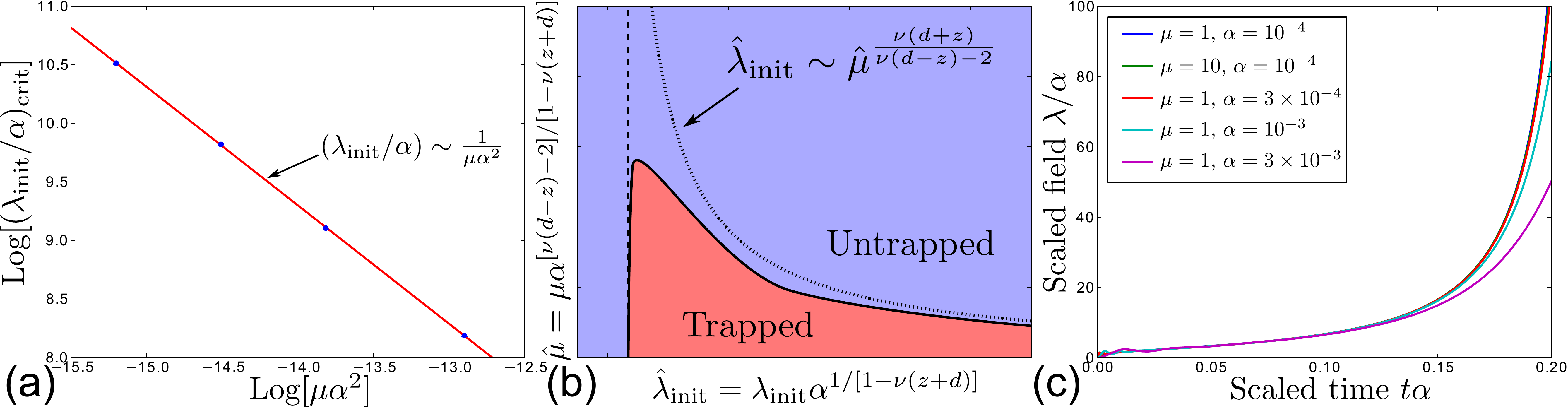}
\caption{Generalized scaling in a linear potential. (a) For the TFI chain with a small slope $\alpha=3\times 10^{-4}$ and small $\hat \mu = \mu \alpha^2$, the critical point for trapping $(\lambda_\init)_\mathrm{crit}$ scales as $1/(\mu \alpha)$ as predicted. (b) Proposed trapping phase diagram for general critical theories in the presence of a linear potential. (c) Scaling collapse at small $\mu \alpha^2$ of the dynamics when the system is initialized at the critical point in its ground state for the TFI chain, showing a lack of trapping.  For $\mu=1$ and the largest values of $\alpha$ shown, deviations are seen at long times due to finite bare mass.}
\label{fig:supp_fig3}
\end{figure}

Now consider a general model with a linear slope. One can play the same games as the previous section to derive a scaling theory in the presence of a slope by nothing that the scaling dimensions of the slope $\alpha$ are $[\alpha] = z + d - 1/\nu$. In order for the low-$\alpha$ scaling limit to be well-defined, we want $\alpha$ to have positive scaling dimension.  This gives a lower critical dimension $d^*_l  = 1/\nu - z$ below which scaling is ill-defined.  Combined with the upper limit $d_u^*= 2/\nu + z$, we require that the dimension $d$ fall within the range.
\be
1/\nu -z < d < 2/\nu + z ~,
\ee
which is clearly the case for both the TFI chain and the Higgs model.

It is then convenient to redefine scaling variables with respect to $\alpha$ instead of $v_\init$, since we want to be able to include the case of $v_\init = 0$. These new scales are given by:
\beq
\nn \ell_{KZ}^{(\alpha)} & =& \alpha^{\nu / (1-\nu z - \nu d)} \\
\nn t_{KZ}^{(\alpha)}  &=& \alpha^{\nu z / (1-\nu z - \nu d)} \\
\nn \lambda_{KZ}^{(\alpha)}  &=& \alpha^{-1 / (1-\nu z - \nu d)} \\
\mu_{KZ}^{(\alpha)}  &=& \alpha (t_{KZ}^{(\alpha)})^2 / \lambda_{KZ}^{(\alpha)} = \alpha^{(2+\nu z - \nu d) / (1 - \nu z - \nu d)} ~,
\label{eq:KZ_alpha}
\eeq
where the $\alpha$ superscript is used to indicate that we rescale with respect to $\alpha$ instead of $v_\init$. In the case of the TFI chain, these reduce to $\ell_{KZ}^{(\alpha)}=t_{KZ}^{(\alpha)}=1/\lambda_{KZ}^{(\alpha)}=\alpha^{-1}$ and $\mu_{KZ}^{(\alpha)}=\alpha^{-2}$. Let us start by considering the case $\mu \gg \mu_{KZ}^{(\alpha)}$ and $\lambda_\init \gg \lambda_{KZ}^{(\alpha)}$, where the initial dynamics are adiabatic. As in the main text, the velocity at the critical point will be
\be
v_c \sim \sqrt{\alpha \lambda_\init / \mu} \sim v_{KZ}^{(\alpha)} \sqrt{\hat \lambda_\init / \hat \mu} ~,
\ee
where 
\be
v_{KZ}^{(\alpha)} = \lambda_{KZ}^{(\alpha)} / t_{KZ}^{(\alpha)}~,~\hat \lambda_\init = \frac{\lambda_\init}{\lambda_{KZ}^{(\alpha)}} ~,~ \hat \mu = \frac{\mu}{\mu_{KZ}^{(\alpha)}} ~.
\ee
As earlier, the excess heat scales as $Q/L^d \sim v_c^{\nu(d+z)/(1+\nu z)}$ and the kinetic energy  is just $K / L^d \sim \alpha \lambda_\init = \alpha \lambda_{KZ}^{(\alpha)} \hat \lambda_\init$, so trapping should occur if $Q/L^d > K/L^d$. This translates to $\hat \lambda_\init < \hat \mu^{\nu(d+z)/(\nu d - \nu z - 2)}$.  For the case of the TFI chain, this prediction reduces to $(\lambda_\init / \alpha) \sim 1/(\mu \alpha^2)$, which we confirm numerically in \fref{fig:supp_fig3}a. Meanwhile, as earlier we expect trapping if $\hat \lambda_\init$ is less than some constant value, yielding a trapping regime:
\be
1 \lae \hat \lambda_\init \lae \hat \mu^{\frac{\nu(d+z)}{\nu d - \nu z - 2}} ~.
\ee
As in the main text, this inequality is more accurately represented as a phase diagram for trapping, which is illustrated in \fref{fig:supp_fig3}b.

The phase diagram represents most of the story, but one question that remains is whether the trapping transition occurs at positive or negative $\hat \lambda_\init$.  It is clear that for $\lambda_\init$ far downhill from the QCP ($\hat \lambda_\init \ll -1$) there cannot be trapping, but we have not found a simple scaling argument to predict whether starting from exactly at the critical point ($\lambda_\init = 0$) will result in trapping. Therefore, we are reduced to simulating this numerically; for example, the case of the TFI chain is shown in \fref{fig:supp_fig3}. We work in the limit $\hat \mu \ll 1$ and small $\alpha$ where we expect the minimal value of $\lambda_\init$ for trapping should occur (see phase diagram). In this limit, bare mass $\mu$ is completely irrelevant since the dressed mass \cite{TestLuca2014_1} scales as $\mu_{KZ}^{(\alpha)}$ and is therefore much larger than the bare mass. Thus the dynamics reduce to single-parameter scaling, as showing in \fref{fig:supp_fig3}c.  For the TFI chain, as in Fig. 3 of the main text, we find no trapping when the system is initialized at the critical, and thus a positive critical value of $\hat \lambda_\init$ for trapping. However, we note that for other models we are not sure what will happen if they are released from their critical points, i.e., we cannot preclude the possibility that some models are trapped in this case. Finally, note that at fixed $\mu$, as $\alpha$ goes to zero, $\hat \mu$ goes to zero. Therefore, starting in the critical regime in the limit of small slope, the bare mass is irrelevant. Thus, except for short time transients, it plays no role in the dynamics, as seen in \fref{fig:supp_fig3}c.

\section{Effect of RG-irrelevant scattering terms}
\label{app:rg_irr_sc_terms}

As argued in the main text, in the presence of RG-irrelevant integrability-breaking terms, the system is expected to thermalize in the long-time limit. Given some initial energy density $E_\init = \alpha \lambda_\init$, here we show that there is a transition in the thermal state between trapping ($E_\init / \alpha^2 > 15.3$) and untrapping ($E_\init / \alpha^2 < 15.3$). Since this argument is based on energy conservation with an \emph{equilbrium} statistical mechanical framework, it does not depend on the mass density $\mu$ of the external field. Combining this with the non-equilibrium phase diagram from earlier, we see that the trapping region remains robust against these thermalizing interactions.

Consider the case of an arbitrary system with a quantum critical point whose low-energy physics is described by an integrable field theory. One example of this is the transverse-field Ising chain, but many other examples exist in one dimension, such as the XXZ model \cite{Yang1966_1}. A standard RG analysis would suggest that near the quantum critical point and at vanishingly small energies, the system is well-described by an integrable theory, meaning that all integrability-breaking interactions are irrelevant. Such RG schemes are well-established in equilibrium, but within the context of Kibble-Zurek scaling, they are also expected to hold \cite{Chandran2012_1,KolodrubetzPRL2012_1}. Our main paper extends this KZ scaling theory to the case of a dynamic field $\lambda$ with a well-defined scaling limit, i.e., $\alpha \to 0$ for the case of a linear potential added to the ground state energy (cf. previous section). However, for any finite $\alpha$, the scaling theory should only hold for a finite amount of time before the interactions start to become important. From an RG sense, at finite but small $\alpha$ we have not yet flowed all the way to the IR, so there remain weak but existent interactions that enable scattering on some long time scale.

Weak interactions perturbing an integrable theory have been studied in many contexts. They are expected to lead to thermalization, as can be seen for example by using second-order perturbation theory to derive an effective Boltzmann equation \cite{StarkArxiv2013_1}. Thus, after a very long time, the state of the spin system should be well approximated by a density matrix $\rho = e^{-\beta H} / Z$, where $H$ is the full interacting Hamiltonian.  This is a difficult density matrix to work with, but fortunately interactions are weak enough that its behavior can in turn be approximated by the ensemble $\rho_0 = e^{-\beta H_0}/Z_0$, where $H_0$ is the integrable part of the Hamiltonian.

\begin{figure}
\includegraphics[width=.35\linewidth]{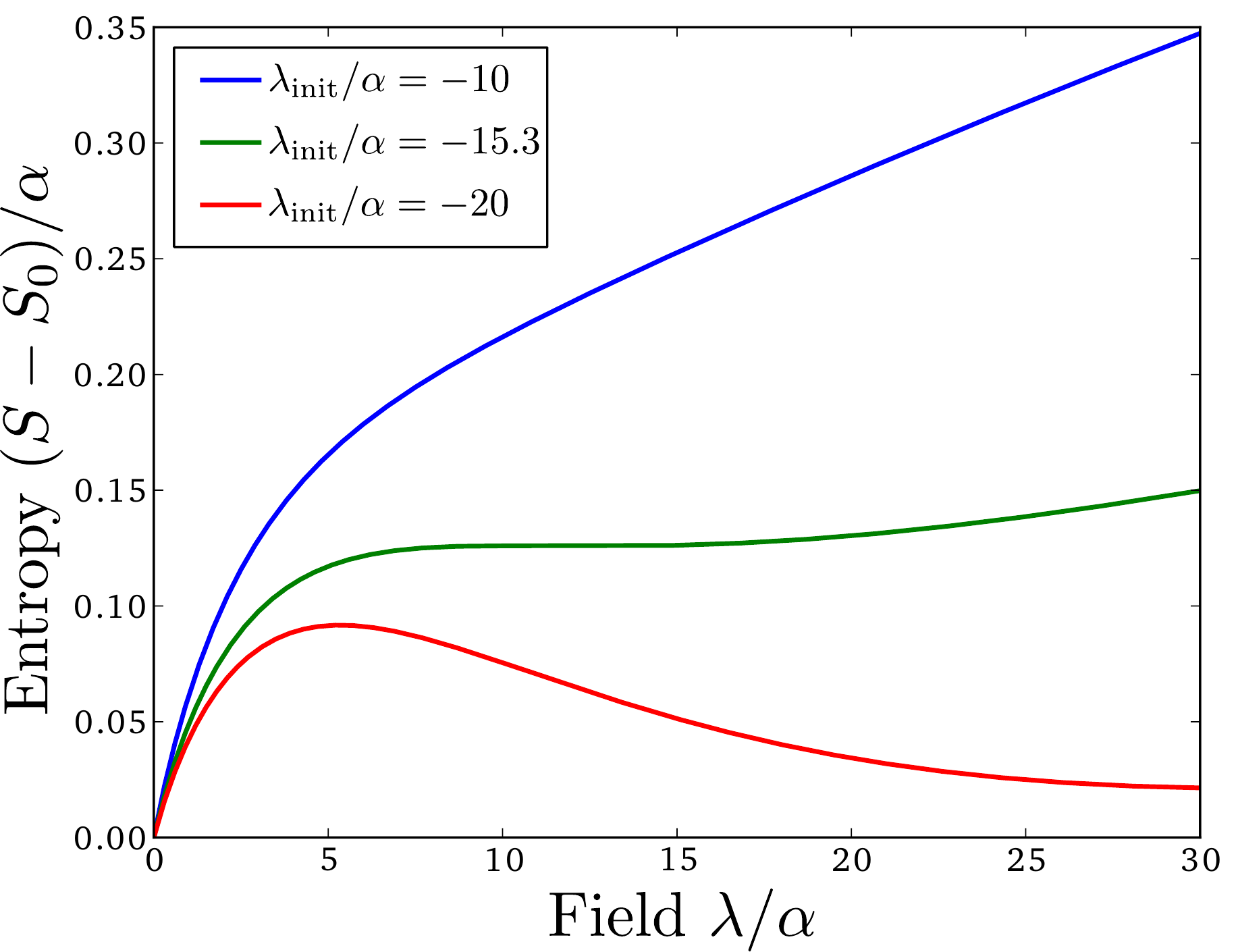}
\caption{Thermal trapping/untrapping transition.  As the initial energy is decreased, the entropy loses its local maximum and the system becomes untrapped. Data is shown for $\lambda_\init/\alpha = -20$ (trapped), $-10$ (untrapped) and $-15.3$ (marginal).}
\label{fig:supp_fig4}
\end{figure}

Taking this approximation to the density matrix, the inverse temperature $\beta(\lambda)$ must be chosen so that the energy of the spins above their ground state matches the loss of potential energy: $\alpha (\lambda_\init - \lambda)$. Within this manifold, the system will equilibrate in the state that (locally) maximizes the entropy by standard principles of statistical mechanics. Whether or not such a maximum exists near the critical point depends on two competing effects. At a constant energy density, the entropy is clearly maximized at the QCP because this is where the most low-lying states exists. However, going towards larger $\lambda$ increases the temperature, and thus the entropy. For the case of the TFI chain, we determine which of these effects wins by numerically determining the entropy $S(\lambda)$. A few characteristic curves are shown in \fref{fig:supp_fig4}a-c, where all the parameters are re-scaled by powers of $\alpha$ to suggest the general form of these curves in the Kibble-Zurek scaling limit. We numerically find that the entropy maximum vanishes at $E_\init / \alpha^2 \approx 15.3$, yielding a thermal trapping/untrapping transition. Remarkably, this transition point is quite close to the one found for the non-equilibrium integrable TFI chain (main text, Fig. 3b), a fact which bears further investigation.

Finally, we note that these ideas can be extended to general models using similar idea as in \sref{sec:supp_scaling_linear}. In many cases, particularly in dimensions higher than one, the expectation is that most quantum critical points will be non-integrable, and thus may thermalize even prior to the inclusion of irrelevant interacting terms \cite{Deutsch1991_1, Srednicki1994_1, Rigol2008_1}. For these cases, the irrelevant terms may modify the trapping/untrapping phase diagrams in small regions around the transition, similar to what they could do for integrable QCPs at finite $\alpha$, but they should not fundamentally alter the shape of the trapped region. If the QCP is non-integrable, however, a similar logic to the above section holds. In particular, assuming that thermalizations occurs to good approximation \emph{within the integrable scaling theory}, then the same Kibble-Zurek scales as in \eref{eq:KZ_alpha} hold. As with the TFI chain, the mass $\mu$ plays no role in thermodynamic trapping, and we are reduced to single-parameter scaling with a transition occuring for sufficiently large initial energy: $\lambda_\init / \lambda_{KZ}^{(\alpha)} \gae 1$. In terms of the general trapping/untrapping phase diagram (\fref{fig:supp_fig3}b), this would correspond to a vertical line such that everything to the right of the line is thermodynamically trapped, and everything to the left is not.  However, we do not currently have an argument for what this precise value of the thermodynamic trapping would be for general integrable theories. Therefore, it is conceivable that reaching thermodynamic equilibrium could decrease the size of the trapping region for some cases.

\end{document}